\documentstyle[11pt,newpasp,twoside]{article}
\def\srm{$\sigma_{\rm RM}$}
\def\rmm{$\langle{\rm RM}\rangle$}
\def\absrmm{$\arrowvert \langle {\rm RM} \rangle  \arrowvert$}
\def\bm{$\langle\mathbf B\rangle$}

\def\edcomment#1{\iffalse\marginpar{\raggedright\sl#1\/}\else\relax\fi}
\reversemarginpar

\begin{document}
\title{Magnetic Fields in Galaxy Clusters}
\author{Federica Govoni}
\affil{Dip. Astronomia, Univ. Bologna, Via Ranzani 1, Bologna, Italy}
\author{Matteo Murgia}
\affil{Ist. di Radioastronomia del CNR, Via Gobetti 101, Bologna,
Italy}

\begin{abstract}
Magnetic fields in galaxy clusters can be investigated using a variety
of techniques. Recent studies including radio halos, 
Inverse Compton hard X-ray emissions and
Faraday rotation measure, are briefly outlined.
A numerical approach for investigating cluster magnetic fields
strength and structure is presented. 
It consists of producing simulated rotation measure, radio halo images,
and radio halo polarization, obtained from 3-dimensional
multi-scale cluster magnetic field models, and comparing 
with observations.
\end{abstract}

\section{Magnetic Field Measures}
A complete description of the astrophysical processes in 
cluster of galaxies requires knowledge of magnetic fields.
The most detailed evidence for this component comes from the 
radio observations:\\
$\bullet$ Some clusters of galaxies exhibit diffuse
non-thermal synchrotron radio halos, associated with the intra-cluster
medium, which extend up to Mega-parsec scales.   
Using minimum energy assumptions, it is possible to estimate
an equipartition magnetic field strength averaged over the
entire halo volume.
These estimates give equipartition magnetic field strengths of
$\simeq$0.1 to 1 $\mu$G (e.g. Bacchi et al. 2003).\\
$\bullet$ In a few cases, clusters containing a radio halo
show an hard X-ray excess emission. 
This emission could be interpreted in terms of
Inverse Compton scattering of the cosmic microwave background photons with the
relativistic electrons responsible for the radio halo emission. In this case,
the measurements of the magnetic field strength 
(e.g. Fusco-Femiano et al. 1999, Rephaeli et al. 1999)
inferred from the ratio of the radio to X-ray luminosities are 
consistent with the equipartition estimates.\\
$\bullet$ Indirect measurements of the magnetic field
intensity can also be determined in conjunction with X-ray observations
of the hot gas, through the study of the Faraday Rotation 
Measure (RM)
of radio sources located inside or behind clusters.
By using a simple analytical approach, magnetic fields of
$\sim$$5-30$ $\mu$G, have been found in cooling flow  
clusters (e.g. Allen et al. 2001; Taylor \& Perley 1993) 
where extremely high RMs have been revealed.
On the other hand, significant magnetic fields have also been 
detected in clusters without cooling flows: the RM measurements
of polarized radio sources through the hot intra-cluster medium leads
to a magnetic field of $2-8$ $\mu$G
which fluctuate on scales as small as $2-15$ kpc.
(e.g. Feretti et al. 1995, Feretti et al. 1999, Clarke et al. 2001,
Govoni et al. 2001, Taylor et al. 2001, Eilek \& Owen 2002). 

The magnetic field strength obtained by RM studies is therefore higher
than the value derived from the radio halo data and from
Inverse-Compton X-ray studies. However, as pointed out by
Carilli \& Taylor (2002) and references therein, all the aforementioned 
techniques are based on several assumptions. 
For example, the observed RMs have been interpreted, until now, in terms
of simple analytical models which consider single-scale magnetic
fields. On the other hand, magneto-hydrodynamic 
cosmological simulations
(Dolag et al. 2002) suggest that cluster magnetic fields may 
span a wide range of spatial scales with a strength that decreases with 
distance from the cluster center.

We developed a numerical approach for investigating the strength and 
structure of cluster magnetic fields. 
It consists of comparing simulated rotation measure, radio halo images,
and radio halo polarization, obtained from 3-dimensional
multi-scale cluster magnetic field models, with observations
(Murgia et al. 2003, submitted).


\section{Simulated Rotation Measures}

The RM is related to the thermal electron density, $n_{\rm e}$, 
and magnetic field along the line-of-sight, $B_{\|}$, 
through the cluster by the equation:
\begin{equation}
{\rm RM} = 812\int\limits_0^L n_{\rm e} B_{\|} {\rm d}l ~~~{\rm rad~m}^{-2}
\end{equation}
where $B_{\|}$ is measured in $\mu$G, $n_{\rm e}$
in cm$^{-3}$ and $L$ is the depth of the screen in kpc.   

We consider a multi-scale magnetic field model with a
three-dimensional power spectrum:
$|B_k|^2 \propto k^{-n}$.
Different power spectrum index will generate different magnetic field 
configurations and therefore will give rise to very different
simulated RM images.

\begin{figure}
\plotfiddle{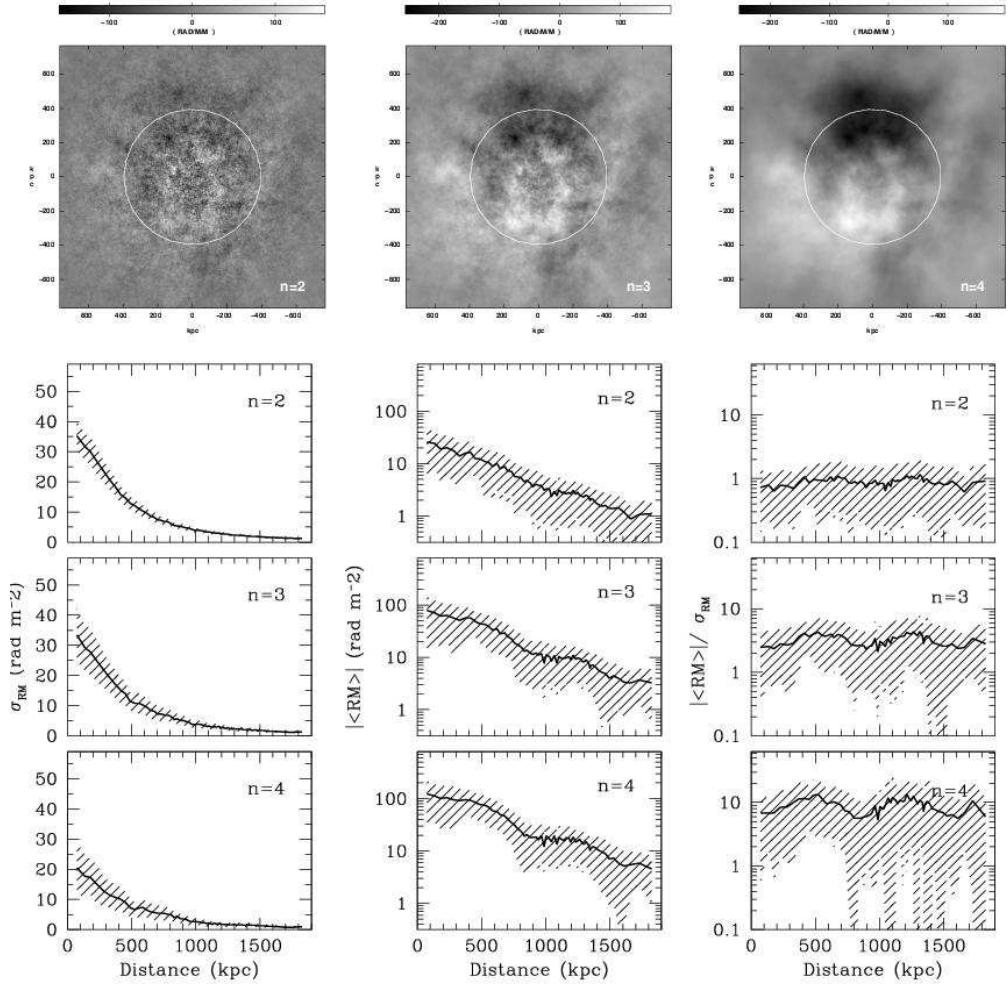}{11cm}{0}{70}{70}{-210}{-10}
\caption{Top: simulated RM images for magnetic 
field power spectrum spectral index $n=2,3,4$. 
The electron gas density of the cluster follow a standard
$\beta$-model with a core radius $r_{\rm c}$=400 kpc 
(indicated by a circle in the figure) a central density 
$n_{\rm e}(0)$=10$^{-3}$ cm$^{-3}$ and $\beta$=0.6.
The three power spectra are normalized to have the 
same total magnetic field energy which is distributed over the range 
of spatial scales from 6 kpc up to 770 kpc. 
The field at the cluster center is \bm$_{\rm0}$ $=1 ~\mu$G and its  
energy density decreases from the cluster center according 
to $B^{2}\propto n_{\rm e}(r)$.
Bottom: radial profiles (\srm, \absrmm~ and \absrmm/\srm~ respectively) 
obtained from the RM simulations described above.
The profiles have been 
obtained by averaging the simulated RM images in regions 
of 50 $\times$ 50 kpc$^2$, which is a typical size for radio galaxies.
} 
\end{figure} 

Fig.~1 (top) shows simulated RM images with different
values of the index $n$ for a typical cluster
of galaxies (see caption for more details).  
RM images, such as those we simulate, cannot be observed for real
clusters of galaxies. However, it is relatively easy to measure the RM
dispersion and mean (\srm~and \rmm) in limited regions by observing
radio sources located at different projected distances from the
cluster center.
Fig.~1 (bottom) shows the simulated profiles of \srm~(left),
\absrmm~(center), and \absrmm/\srm~(right), as a function of the 
distance from the cluster center. While both \srm~and \absrmm~increase 
linearly with the cluster magnetic field strength,  
the ratio \absrmm/\srm~ depends only on the magnetic field power spectrum slope
for a given range of fluctuation scales.
This means that the comparison of RM data of radiogalaxies
embedded in a cluster of galaxies with simulated profiles, can infer 
the strength and the power spectrum slope of the
cluster magnetic field. 
The comparison of our simulations with data (Murgia et al. 2003,
submitted), indicates magnetic fields 
strength a factor of about 2 lower than that predicted by the 
single-scale magnetic field approximation widely used in literature
and a rather flat spectral index $n \simeq$ 2.

\section{Simulated Radio Halo Images and Radio Halo Polarization}

\begin{figure}
\plotfiddle{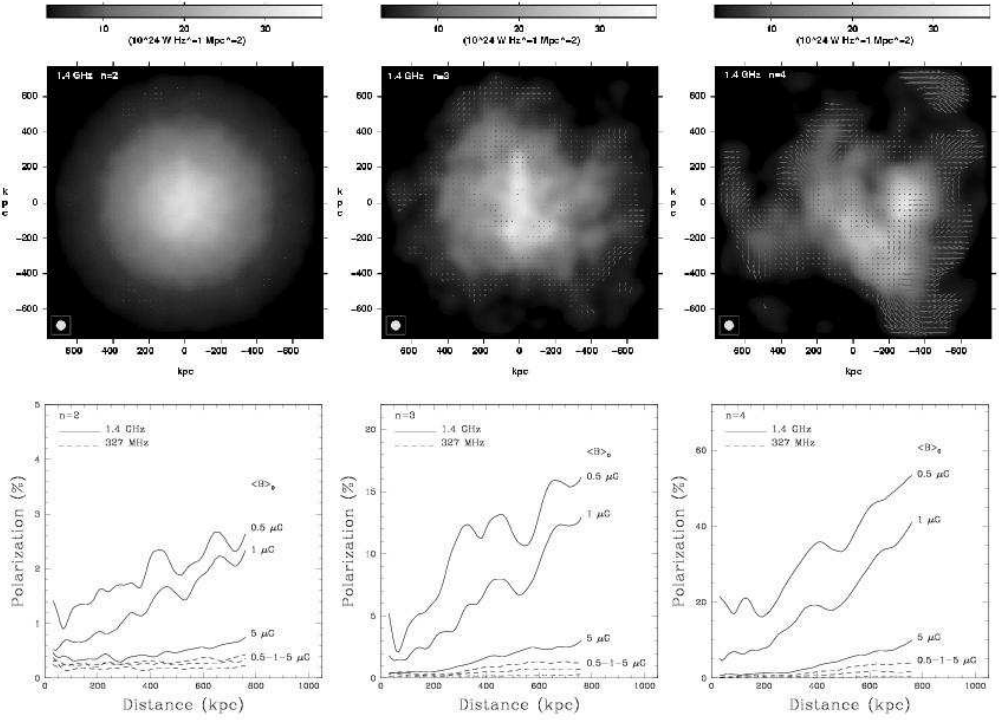}{9cm}{0}{70}{70}{-215}{-15}
\caption{Simulated halo brightness and polarization for cluster at a
 distance of $z=0.05$ as it would be observed with a beam of 45\arcsec~($H_{\rm 0}=50 ~\rm km\,s^{-1}Mpc^{-1}$).
Top: simulated halo images at 1.4 GHz for different values of the 
 magnetic field power spectrum slope $n$ and $\langle B\rangle_{\rm 0}=$1 
${\rm \mu G}$; the vectors lengths are
proportional to the degree of polarization, with 100 percent corresponding
 to 100 kpc on the sky.  Field directions are those of the E-vector. 
Bottom: radially averaged 
 profiles of the polarization percentage at 327 MHz and 1.4 GHz for 
three values of the magnetic field 
strength, namely  $\langle B\rangle_{\rm 0}=$0.5, 1 and 5 ${\rm \mu G}$. 
}
\end{figure}

Radio halos observations may provides 
important information about the cluster magnetic field structure
since different values of the power spectrum spectral
index will generate very different total intensity and polarization
brightness distributions for the radio halo emission.

So far, polarization emission from radio halos has not been
detected. The current upper limits to the
polarization at 1.4 GHz, for beams of
about 45\arcsec, are a few percent ($3-4$\%).

We simulated the expected total intensity and polarization brightness
distribution at 1.4 GHz and 
327 MHz, as it would be observed with a beam of 45\arcsec, 
for different values of the magnetic field strength and 
power spectrum index, by introducing in
the 3-dimensional magnetic field an isotropic population of
relativistic electrons.

Fig. 2 (top) shows simulated radio halo brightness and 
polarization percentage distributions at 1.4 GHz 
(see caption for more details).
Fig. 2 (bottom) shows the expected fractional 
polarization profiles at 1.4 GHz and 327 MHz for the different values of the average magnetic field strength and power spectrum spectral index. Our results indicate that a power spectrum slope steeper than $n=3$ and
 a magnetic field strength lower than $\sim 1 {\rm \mu G}$
 result in a radio halo polarization percentage at a frequency of 1.4 GHz
that is far in excess of the current observational upper limits at 45\arcsec~
resolution.
This means that, in agreement with the RM simulations, 
either the power spectrum spectral index is flatter
than $n=3$ or the magnetic field strength is significantly higher than
$\sim 1 {\rm \mu G}$. The halo depolarization at 327 MHz is 
particularly severe and the expected polarization percentage
at this frequency is always below 1\%. 
Moreover we also found that the magnetic field power
spectrum slope has a significant effect in shaping the radio halo. In
particular, flat power spectrum indexes ($n<3$) give raise to
very smooth radio brightness images (under the assumption that the
radiating electrons are uniformly distributed). 


\section{Conclusions}
The numerical approach presented here demonstrates how 
the dispersion and mean of the 
RM measured in radio galaxies embedded in a cluster of galaxies
can be used to constrain not only the strength but also the
power spectrum slope of the intra-cluster magnetic fields. 
Moreover, the study of the polarization properties 
of a large scale radio halo, if it is present in a cluster,
can be used to improve the estimates based on the RM analysis.\\

\noindent
{\it Acknowledgment.}~ We thank L.Feretti, G.Giovannini,
 D.Dallacasa, R.Fanti, G.B. Taylor and K.Dolag for their collaboration.

\end{document}